\documentclass[12pt]{iopart}

\usepackage{graphicx}
\begin{document}

\title[]{Traffic-driven SIR epidemic model on networks}

\author{Cunlai Pu, Siyuan Li and Jian Yang}
\address{School of Computer Science and Engineering, Nanjing University of Science and Technology, Nanjing 210094,  China
}
\ead{pucunlai@njust.edu.cn}
\begin{abstract}
We propose a novel SIR epidemic model which is driven by the transmission of infection packets in networks. Specifically, infected nodes generate and deliver infection packets causing the spread of the epidemic, while recovered nodes block the delivery of infection packets, and this inhibits the epidemic spreading. The efficient routing protocol governed by a control parameter $\alpha$ is used in the packet transmission. We obtain the maximum instantaneous population of infected nodes, the maximum population of ever infected nodes, as well as the corresponding optimal $\alpha$ through simulation. We find that generally more balanced load distribution leads to more intense and wide spread of an epidemic in  networks. Increasing either average node degree or homogeneity of degree distribution will facilitate epidemic spreading. When packet generation rate $\rho$ is small, increasing $\rho$ favors epidemic spreading. However, when $\rho$ is large enough, traffic congestion appears which inhibits epidemic spreading. 
\end{abstract}

\pacs{89.75.Hc, 89.75.Fb, 05.10.-a}
\maketitle
\section{Introduction}
As the increase of connectivity in and between different complex systems, spread of many diseases or viruses is becoming more and more prevalent in our society\cite{Dor08,Barrat,Newman09,pastor2014}.
For instance, outbreaks of many infectious diseases, including Severe Acute Respiratory Syndromes (SARS), Swine flu (H1N1), and the recent Ebola virus, caused great damage and loss of life. The spread of computer and mobile phone viruses brought about a great deal trouble to human life and serious damage to economy. Understanding the intrinsic mechanisms of those spreading processes and designing efficient control strategies become very important and urgent tasks, which bring together a lot of researchers from areas of biology, sociology, mathematics, physics, engineering, etc\cite{Newman09}. 

Mathematical modeling of epidemic spreading has a long history of more than two hundred years\cite{pastor2014}. Generally, the population is divided into several classes: susceptible, infected and recovered individuals. Susceptible individuals represent those who can contract the infection. Infected individuals were previously susceptible individuals and got infected by the disease. Recovered individuals are those who have recovered from the infection. In the susceptible-infected-susceptible (SIS) model\cite{pastor2014}, infected individuals can recover from the disease and become susceptible individuals again. While in the susceptible-infected-recovered (SIR) model\cite{pastor2014}, infected individuals no longer get infected after recovery from the disease, which are assumed to get the permanent immunity. In classical epidemiology, a common assumption is that individuals in a class is treated similarly, and have equal probability to contact with everyone else\cite{Newman09,pastor2014}. However, the recent abundance of data demonstrates that both the connectivity pattern and the contact rate are heterogeneous among real-world complex networks\cite{Newman09,barabasi09}, which means the traditional deterministic differential equations and many other related results of epidemic processes are inadequate in real-world situations. This great stimulates the research of epidemics on real-world complex networks\cite{pastor2014}. Due to the complexity of real-world networks, the mean-field approach\cite{pastor01,Yangz12,sahneh13} and the generating function approach\cite{MEJNEWMAN02} are used to drive the analytic results of epidemics spreading. One of the remarkable results obtained by Pastor-Satorras and Vespignani\cite{pastor2014, pastor01} shows that in the limit of a network of infinite size, the epidemic threshold of the SIS model tends to zero asymptotically in scale-free networks with power-law parameter in (2, 3]. For SIR model, it was found that in the thermodynamic limit, not only the threshold tends to vanish, but also the time for the stabilization of the infection becomes very small\cite{barth04,barth05}. By using the message-passing approach, Karrer and Newman\cite{karrer10} calculated the probabilities for any node and any time to be in state S, I, and R on tree structure. Many other explicit results of SIR model are obtained by mapping the SIR model to the percolation process\cite{MEJNEWMAN02,kenah07,miller07}. Also, effects of degree correlations\cite{goltsev08,goltsev12}, clustering\cite{serrano06,miler09}, weights and directions of edges\cite{Yangz12,gang05,chu11} on epidemic spreading are broadly discussed. On the other hand, various efficient immunization protocols\cite{pastor2014,PASTOR02,van2012}  have been designed for controlling the spread of epidemics on networks. Recently much attention has been transferred to epidemic spreading in temporal and multiplex networks\cite{starnini,Granell,zhao14}.

Addition to diseases or viruses, there are usually many other substances spreading in networks like information packets, goods, ideas, etc., which depend on the specific types of the networks. Epidemic spreading is often coupled with the delivery of these substances. For example, HIV spreads through the exchange of body fluids among individuals in contact networks. Computer viruses spread with the delivery of information packets in computer networks. Flu often spreads by air traffics among different spatial areas. Therefore, understanding the mechanisms of these coupled spreading processes and how these processes affect each other is significant for designing efficient epidemic immunization strategies. Meloni et al\cite{meloni} first studied the effects of traffic flow on epidemic spreading. They found that the epidemic threshold in the SIS model decreases as flow increases, and emergence of traffic congestion slows down the spread of epidemics. Then, Yang et al\cite{yang11,yang14} further studied the relation between traffic dynamics and the SIS epidemic model, and found that the epidemic can be controlled by fine tuning the local or global routing schemes. Furthermore, they obtained that the epidemic threshold can be enhanced by cutting some specific edges in the network\cite{yangh13}. The impacts of traffic dynamics on SIR epidemic model haven't been reported in literature. In this paper, we study the traffic-driven SIR spreading dynamics in complex networks. We focus on the instantaneous size of infected population, and the final size of ever infected population. Based on these two properties, we study how the packets transmission process governed by given routing protocols affects the epidemic spreading.

\section{Model}
Our model includes two coupled processes: packet delivery process and the epidemic spreading process. We will introduce our model in the context of computer networks.
\subsection{packet delivery process}
We assume that nodes in the network are identical which can generate, receive and deliver information packets.  Each node has a queue obeying the First-In-First-Out (FIFO) rule for storing packets. The length of the queue is set infinite. Load of a node is the number of packets in its queue. Every node generates packets at a rate $\rho$. For example if $\rho=1.5$, each time an arbitrary node generates one packet definitely and another one with probability 0.5. The destination nodes of the packets are chosen randomly, and the packets will be removed from the network after arriving at the destination nodes. The transmission of packets is governed by the efficient routing protocol proposed by Yan et al\cite{yan06}. For an arbitrary path $p$ of length $l$ between node $s$ and $d$, denoted as $\langle s, n_1, n_2, \cdots, n_{l-1}, d \rangle$, its routing cost is defined as follows:
\begin{equation}
\phi(p)=\sum_{i=1}^{l-1}k_{i}^{\alpha}.
\end{equation}
Where $k_i$ is the degree of node $i$, and $\alpha$ is a control parameter. The sum runs over all the intermediate nodes of path $p$. The efficient paths for delivering packets are defined to be those which have the minimum routing costs.  If there are many efficient paths between two nodes, we randomly chose one for delivering packets. According to Eq. 1, $\alpha$ determines the routing cost of a path. When $\alpha>0$, paths with large-degree nodes usually have large routing costs. Thus, efficient paths tend to be those paths composed of small-degree nodes when $\alpha>0$, and vice versa. Each time a node can deliver at most $C$ packets. When $C=\infty$, all the packets can be delivered without delay, there is no traffic congestion. The overall load of the network is constant after a short transient time. When $C$ is a constant value, there is a critical packet generation rate $\rho_{c}$. When $\rho<\rho_c$, there is no traffic congestion. When $\rho>\rho_c$, the network generates more packets than it can deliver. As a result, the overall load of the network increases with time, which is the traffic congestion phenomenon.
We use the order parameter to characterize the traffic congestion, which is as follows\cite{arenas01}:
\begin{equation}
\psi=\frac{1}{\rho N}\lim\limits_{t\to\infty}\frac{\langle w(t+\Delta t)-w(t)\rangle}{\Delta t}.
\end{equation}
Where $w(t)$ is the total number of packets in the network at time $t$. When $\psi=0$, the generated and delivered packets are balanced, and the network is under the free flow state. When $\psi>0$, packets accumulate continuously in the network, which indicates that there exists traffic congestion.
\subsection{Epidemic spreading process}
As in the traditional SIR model, nodes in the network are divided into three classes: susceptible nodes, infected nodes and recovered nodes. Initially, all the nodes in the network are susceptible nodes, which perform the normal function of generating, delivering, and receiving packets, and the network flow is stable. 
Then we randomly select a node to be an infected one which is the original source of the infection. Infected nodes generate infection packets instead of normal packets at each time step. With the delivery of these infection packets, more and more susceptible nodes get infected after receiving the infection packets. Infected nodes get recovered and become recovered nodes with probability $\mu$ at each time step. Recovered nodes generate normal packets, and they can also make the infection packets into normal packets. Thus, recovered nodes block the epidemic spreading. The epidemic spreading ends when all the infected nodes become recovered nodes. The transitions between the susceptible, infected, and recovered nodes for our model are shown in figure 1.

\begin{figure}
 \centering
\includegraphics[width=4in,height=1in]{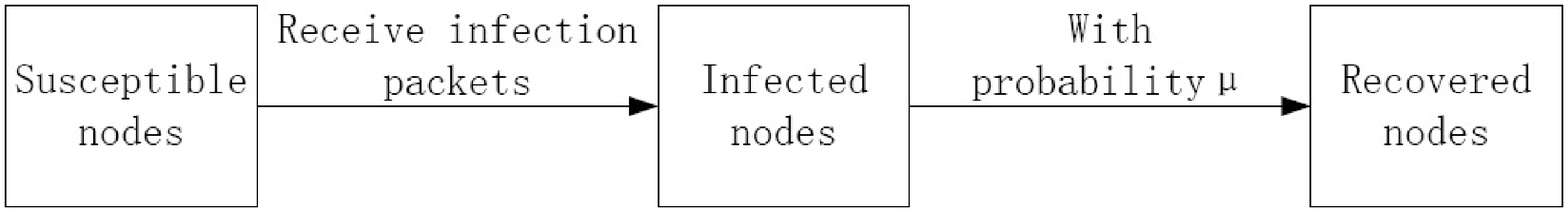}
\caption{Transitions of susceptible nodes, infected nodes and removed nodes. }
\end{figure}
 \section{Results}
The underlying networks are random networks generated by the Erd\"{o}s-R\'{e}nyi (ER) model\cite{Erdos} and scale-free networks generated by the static model\cite{Goh01}. Also, some real-world networks\cite{guimera,shiwei,damic05} are used in the simulations. 
\subsection{Evolution of our model}
We assume that, at time $t$, the numbers of susceptible nodes, infected nodes and recovered nodes are $S(t)$, $I(t)$ and $R(t)$ respectively.
First, we study the time evolution of our model on the ER model, the static model, and the Email network\cite{guimera}. We add the infection to the network at $t=150$ when the network is under free flow state, by randomly selecting a node to become infected. In figure 2 (a), (b) and (c), we see that $S(t)$ decreases with $t$ greatly, and then tends to be stable. On the contrary, $R(t)$ increases with $t$ abruptly, and then converges at $R_e $, which is the maximum number of recovered (or ever infected) node. $R_e$ reflects the range of the infection.
Differently from $S(t)$ and $R(t)$, $I(t)$ increases with $t$ first, then decreases with $t$. The peak of $I(t)$,  denoted by $I_p$,  represents the maximum instantaneous number of infected nodes during the epidemic spreading process. $I_p$ reflects how intense the epidemic spreading is.  At any time $t$, the sum of $S(t)$, $I(t)$, and $R(t)$ equals the size of the network.
\begin{figure}
 \centering
\includegraphics[width=4in,height=3in]{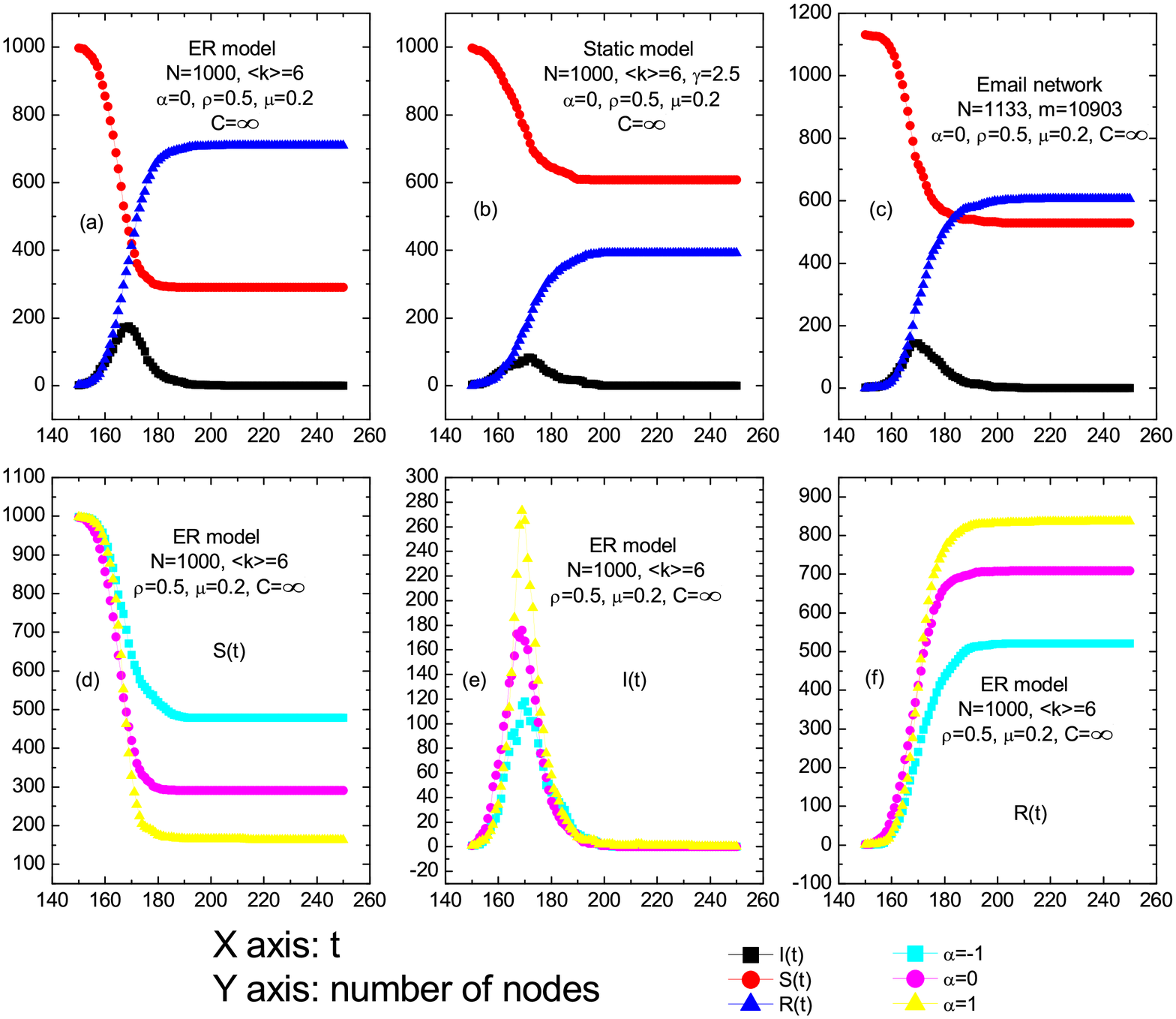}
\caption{Time evolution of our  model on the networks }
\end{figure}
\subsection{Impacts of $\alpha$}
Routing parameter $\alpha$ is one of the important factors in our model. $\alpha$ determines the efficient paths for the delivery of normal and infection packets. In figure 2(d), (e), and (f), we see that $S(t)$, $I(t)$ and $R(t)$ all vary with $\alpha$, and for any of the three, the trends of the curves of different $\alpha$ are similar.
We focus on the maximum instantaneous size of infected population $I_p$, and the final size of ever infected population $R_e$. 
In figure 3 (a), (b) and (c), we see that both $I_p$ and $R_e$ increase with $\alpha$ first, then decrease with $\alpha$. There are optimal $\alpha$ which correspond to the maximum $I_p$ and $R_e$ respectively. Note that the optimal $\alpha$ for $I_p$ and $R_e$ are close, but  not necessarily the same. There are jumps in both $I_p$ and $R_e$ when $ \alpha$ is near zero. To explain these results, we calculate the load variance $ \sigma_L$ of nodes, which is defined as follows:
\begin{equation}\label{pb1}
\left \{
\begin{array}{rl}
\overline{L_i}&= \frac{1}{T}\sum_{t=0}^{T}L_{i}(t),\\
\sigma_L &= \sqrt{\frac{\sum_{i=1}^{N}(\overline{L_i}-\frac{1}{N}\sum_{j=1}^{N}\overline{L_j})^2}{N}}.\\
\end{array}
\right.
\end{equation}
Where $L_{i}(t)$ is the load of node $i$ at time $t$. $T$ is a constant value and is large enough to ensure accurate calculation of the average load $\overline{L_i}$ of node $i$. When $\sigma_L$ is small, the load distribution is relatively even in the network, and vice versa.
In figure 3 (d), (e) and (f), $\sigma_L$ decreases first and then increases with $\alpha$. There is optimal $\alpha$ which leads to the minimum $\sigma_L$. Interestingly, the values of the optimal $\alpha$ for $\sigma_L$, $I_p$ and $R_e$ are close, which indicates that homogeneous load distribution facilitates the epidemic spreading. There is also abrupt decrease in $\sigma_L$ when $\alpha$ is near zero. This is because the efficient paths are very different for $\alpha < 0$ and $\alpha >0$, and the load is abruptly redistributed from large-degree nodes to small-degree nodes when $\alpha$ increases from below zero to above zero. This also accounts for the jumps in $I_p$ and $R_e$. The results are consistent for both the models networks and the Email network as shown in figure 3.
\begin{figure}
 \centering
\includegraphics[width=4in,height=3in]{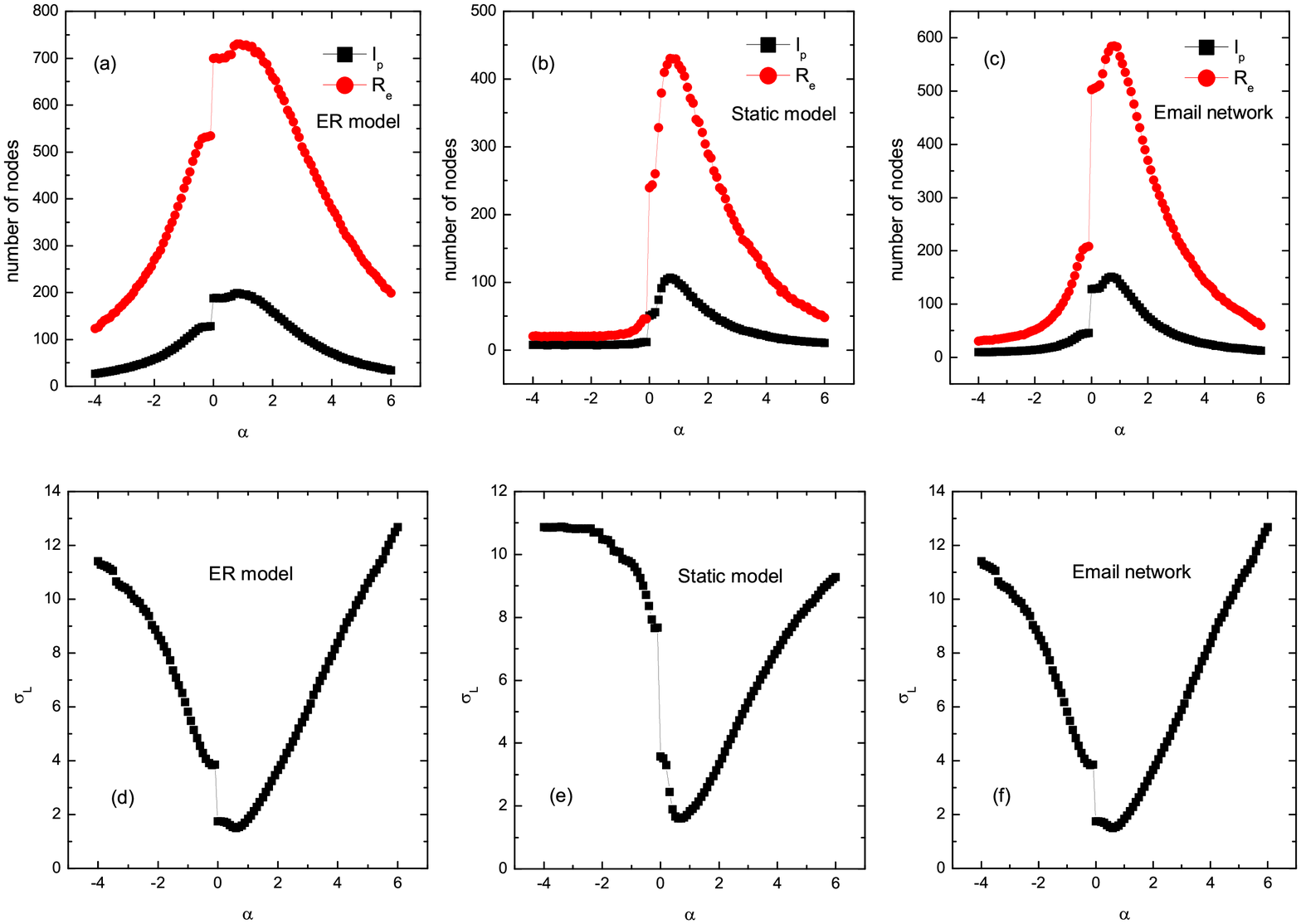}
\caption{$I_p$, $R_e$ and $\sigma_L$ vs. $\alpha$. The network models,   the Email network, as well as all the parameters  are the same as in figure 2. The results are the average over 100 independent runs. }
\end{figure}
\subsection{Impacts of $\rho$}
Packet generation rate $\rho$ also has great impacts on the epidemic spreading. In figure 4, we see that the peak of $ R_e$ for $\rho=0.5$ is almost 4 times of the peak of $R_e$ for $\rho=0.2$. The peak of $I_p$ for $\rho=0.5$ is almost 6 times of the peak of $I_p$ for $\rho=0.2$. Also the positions of the peaks for $\rho=0.1, 0.2$ and 0.5 are different. In figure 5 (a), we clearly see both the maximum $I_p$ and the maximum $R_e$ increase with $\rho$. When $\rho$ becomes large, infected nodes will generate more infected packets, which facilitates the epidemic spreading.  In figure 5 (b), we see that $\alpha_{opt}$ decreases with $\rho$, which indicates an increase of dependency on large-degree nodes.  To illustrate this, we focus on the top 1\% largest degree nodes, and calculate the average ratio $\eta$ of the number of infection packets a large degree node delivered when it was infected to the number of infection packets it cured after it recovered.
In figure 6, $\eta$ increases with $\rho$, which means that the role large-degree nodes play in facilitating the epidemic spreading becomes more and more remarkable compared to the role they play in inhibiting the epidemic spreading.
This explains why the epidemic spreading becomes more dependent on large-degree nodes to spread widely when $\rho$ increases as shown in figure 5 (b). 
\begin{figure}
 \centering
\includegraphics[width=5in,height=2in]{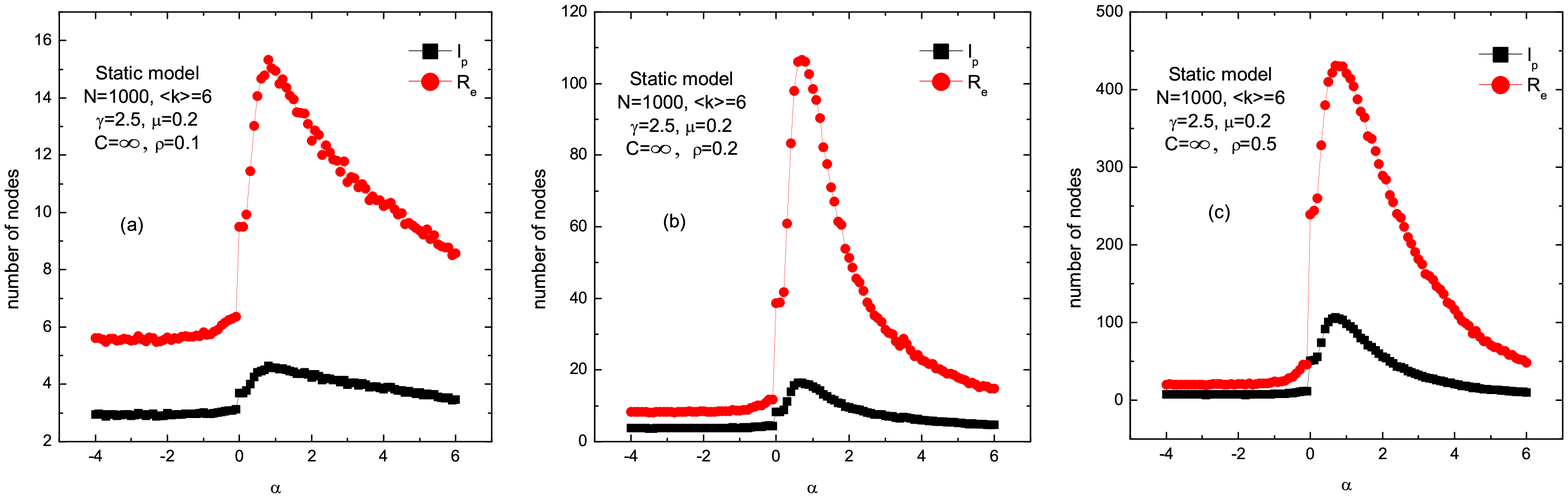}
\caption{$I_p$ and $R_e$  vs. $\alpha$ for different $\rho$. The results are the average over 100 independent runs. }
\end{figure}
\begin{figure}
 \centering
\includegraphics[width=4in,height=2in]{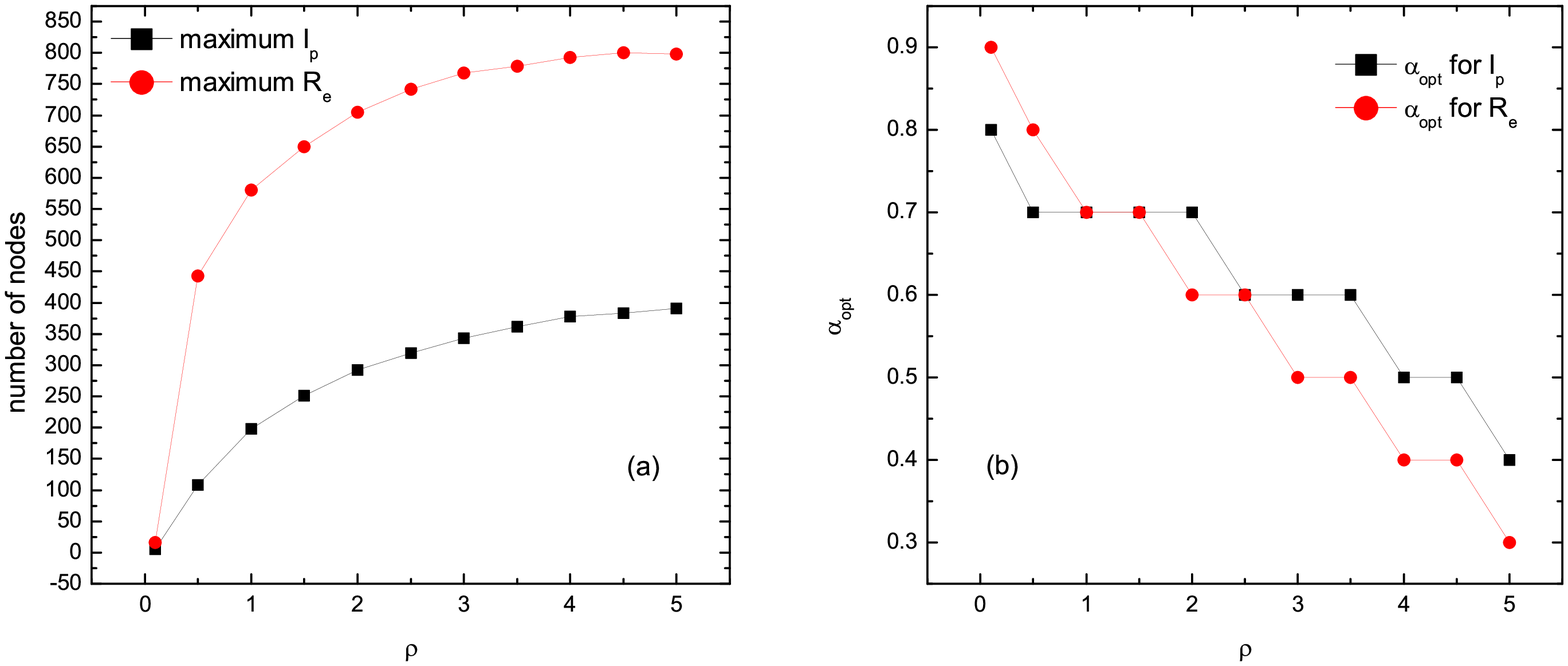}
\caption{The maximum $I_p$, the maximum $R_e$ and the corresponding $\alpha_{opt}$ vs.  $\rho$. The network model and the parameters are the same as in figure 4.  The results are the average over 100 independent runs. }
\end{figure}
\begin{figure}
 \centering
\includegraphics[width=3in,height=3in]{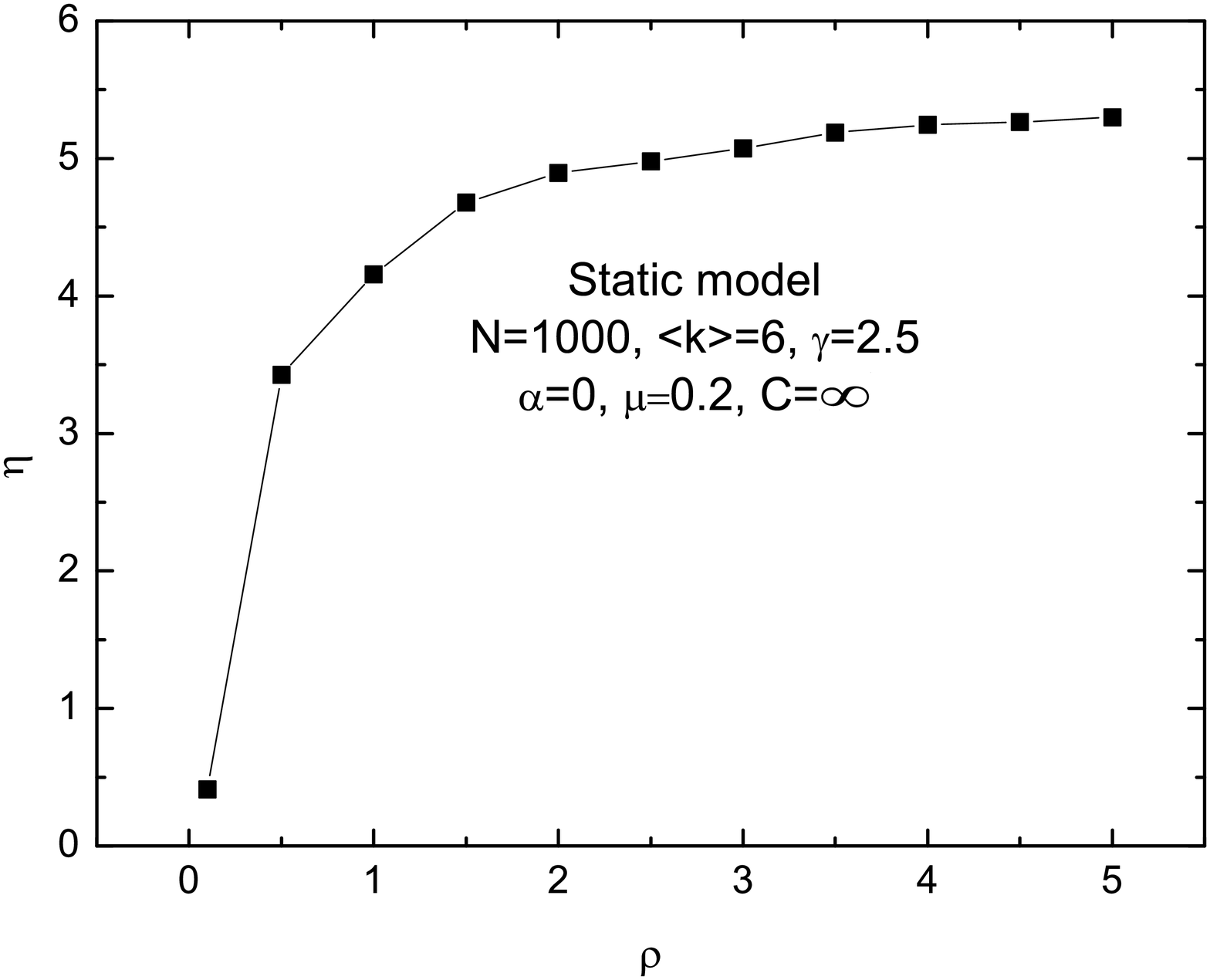}
\caption{The ratio $\eta$  vs. $\rho$.  The results are the average over 10 independent runs. }
\end{figure}
\subsection{Impacts of network structure}
First, we show the impacts of network density on the epidemic spreading. In figure  7 (a) and (c), we see that the maximum $I_p$ and the maximum $R_e$ increase with average degree $\langle k \rangle$ for both random networks and scale-free networks. This indicates that more edges facilitate the epidemic spreading. In figure  7 (b) and (d), $\alpha_{opt}$ decreases with $\langle k \rangle$, but is large than zero. This means that, to realize an intense and wide epidemic spreading, generally the paths for the delivery of infection packets should be biased towards including small-degree nodes, but when the network becomes dense, the degree of the dependence of large-degree nodes in the transmission of infection packets should be increased accordingly.
\begin{figure}
\centering
\includegraphics[width=4in,height=3in]{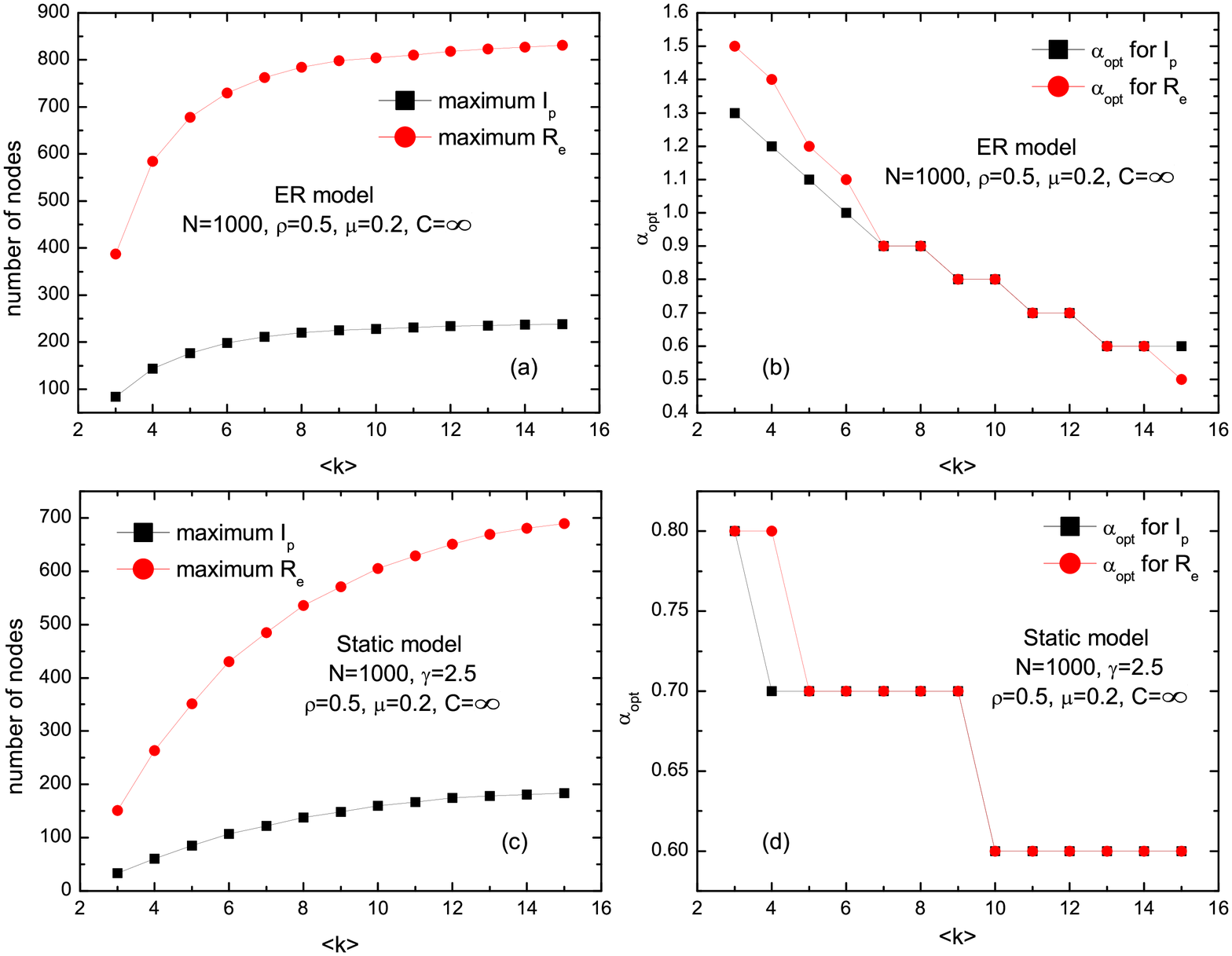}
\caption{The maximum $I_p$, the maximum $R_e$ and the corresponding $\alpha_{opt}$ vs. $\langle k \rangle$. The results are the average over 100 independent runs. }
\end{figure}
Then we show the impacts of power-law parameter $\gamma$ on the epidemic spreading in figure 8. Both the maximum $I_p$ and the maximum $R_e$ increase with $\gamma$ first, and then tend to be stable. The $\alpha_{opt}$ increases slightly with $\gamma$. We infer from figure  8 that homogeneous network structure facilitates the epidemic spreading. 
\begin{figure}
\centering
\includegraphics[width=4in,height=2in]{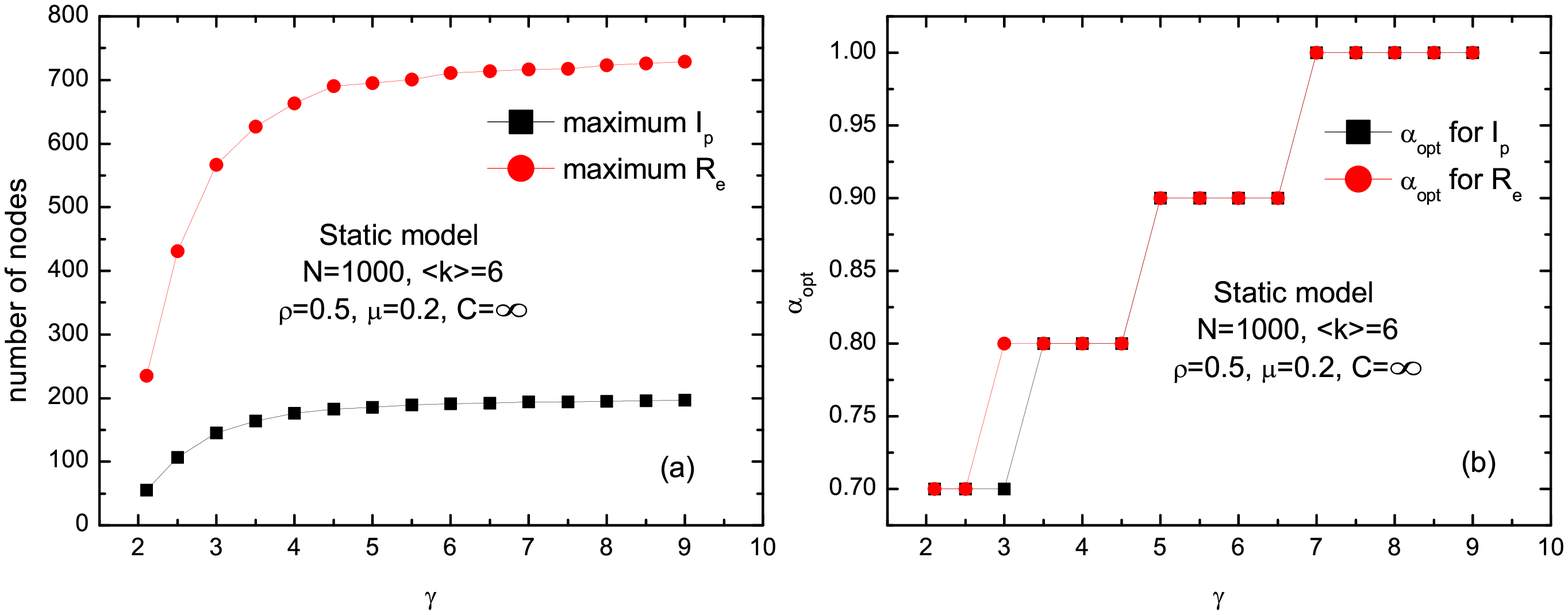}
\caption{The maximum $I_p$, the maximum $R_e$ and $\alpha_{opt}$ vs. $\gamma$. The results are the average over 100 independent runs. }
\end{figure}
\subsection{Impacts of traffic congestion}
When delivery capacity $ C$ is a constant value, and the packet generation rate $\rho$ is large enough, the packets can't be delivered in time, and then the number of packets accumulated in the network increases with time, which is the traffic congestion phenomenon. In the simulation, $C$ is set to 10, $\psi$ is calculated according to Eq. 2 to quantify the degree of traffic congestion. In figure  9 (a), we see that only when $\alpha$ is in [0.6, 1], there is no traffic congestion in the network, which is indicated by $\psi=0$. Otherwise, there exists traffic congestion where $\psi >0$. In figure  9 (b), we see that both $I_p$ and $R_e$ increase with $\alpha$ first, and then decrease with $\alpha$. The optimal $\alpha$ for $I_p$ and $R_e$ are 0.7 and 0.9 respectively, where there is no traffic congestion. Also, we obtain that when the traffic congestion is not serious like $\alpha$ in [1.5, 2] (figure 9 (a)), the epidemic spreading can still spread intensely and widely, which is inferred by $I_p$ and $R_e$ in figure  9 (b).
\begin{figure}
\centering
\includegraphics[width=4in,height=2in]{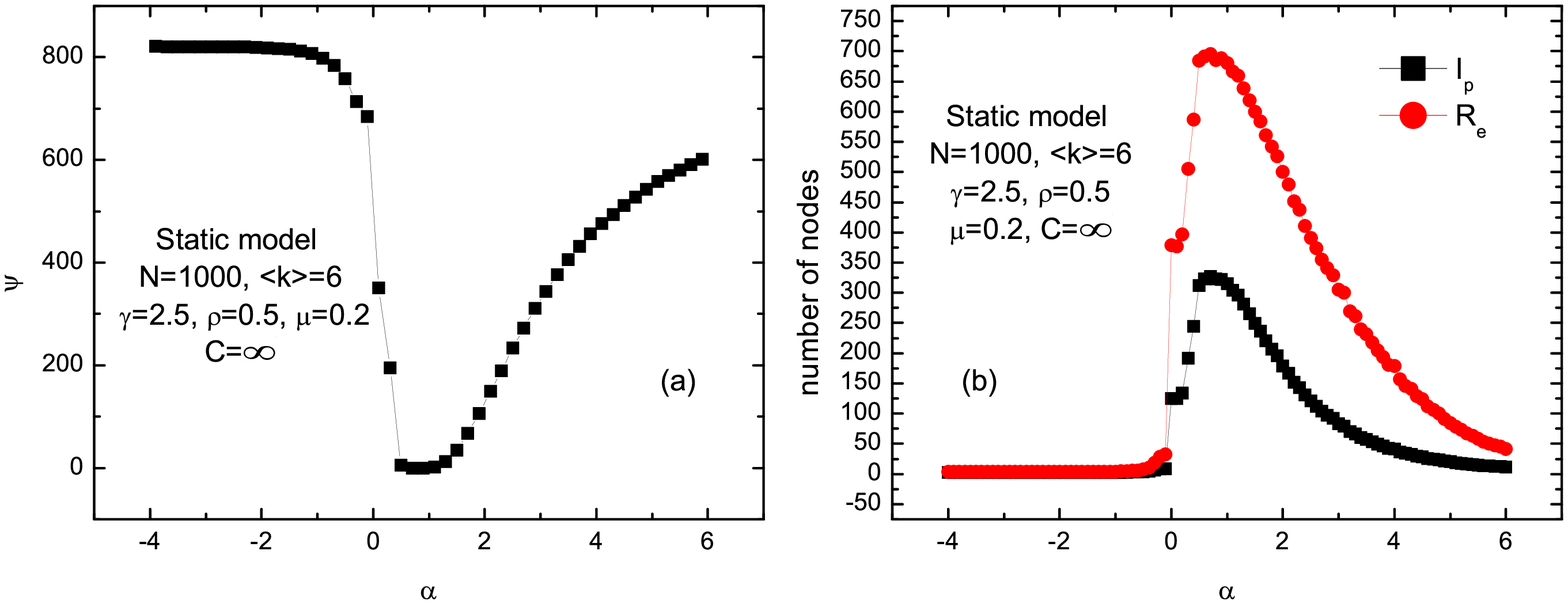}
\caption{$\psi$, $I_p$ and $R_e$ vs. $\alpha$. The results are the average over 100 independent runs. }
\end{figure}
Then we fix $\alpha$ to be zero, and study how the maximum $I_p$ and the maximum $R_e$ vary with $\rho$. In figure  10, we obtain that both the maximum $I_p$ and the maximum $R_e$ increase first and then decrease with $\rho$, which is consistent for both random networks and scale-free networks. The reason for these results is that when $\rho$ is small, there is no traffic congestion, and the infection becomes intense and wide spread with increase of $\rho$. However, when $\rho$ is large enough, traffic congestion appears, which inhibits the epidemic spreading. 
\begin{figure}
\centering
\includegraphics[width=4in,height=2in]{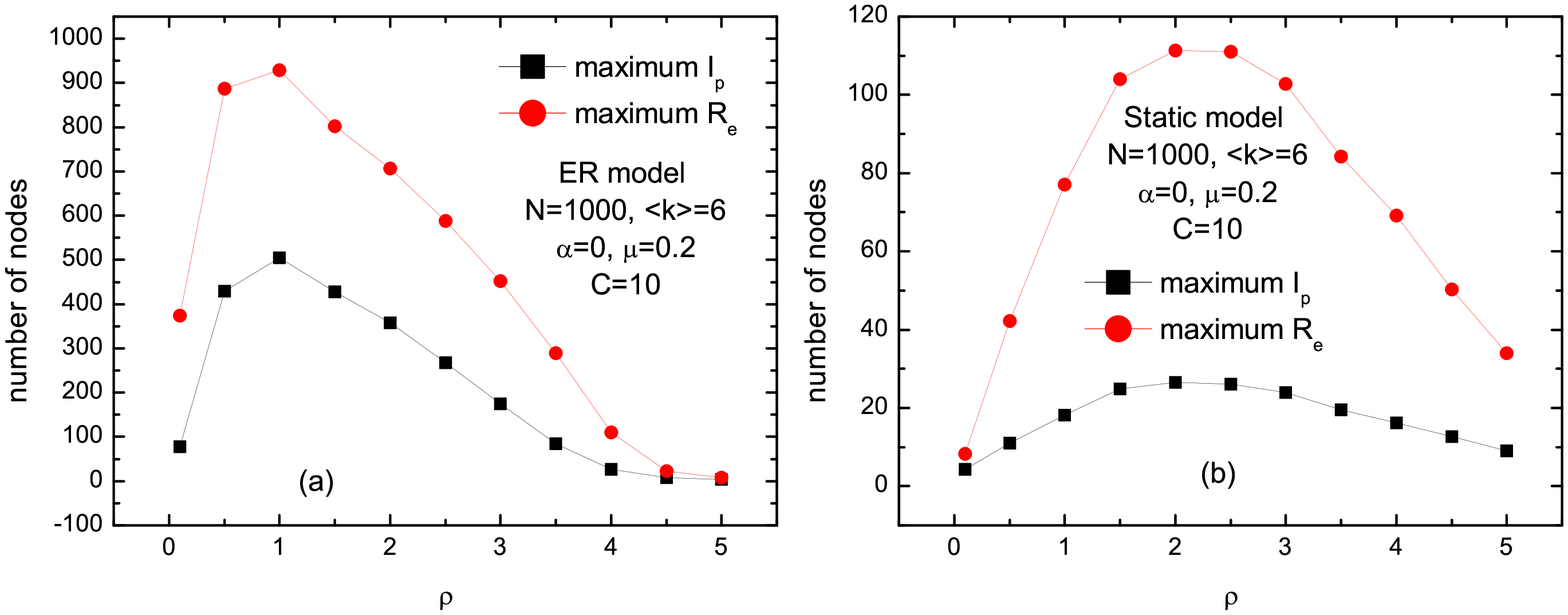}
\caption{The maximum $I_p$ and the maximum $R_e$ vs. $\rho$. The results are the average over 100 independent runs. }
\end{figure}
\subsection{Impacts of routing protocols}
In addition to the efficient routing protocol, we also study the impacts of the other static routing protocols on the epidemic spreading.
If $k$ is replaced with $\lg k$ in Eq. 1, then we get the cost function of the optimal routing protocol\cite{Wangk11}. For the optimal routing protocol, we only present the results of $ I_p$ and $R_e$ vs. $\alpha$ on three real-world networks\cite{guimera,shiwei,damic05} in figure  11. There are also optimal $\alpha$ which correspond to the maximum $I_p$ and the maximum $R_e$ respectively. When $\alpha$ is near zero, there are also jumps in $I_p$ and $R_e$ due to the significant change of paths for delivering packets. For the optimal routing and the efficient routing, either the maximum $I_p$ or the maximum $R_e$ is very close. The difference lies in that $I_p$ and $R_e$ of the optimal routing vary much slower than that of the efficient routing.
\begin{figure}
 \centering
\includegraphics[width=5in,height=2.5in]{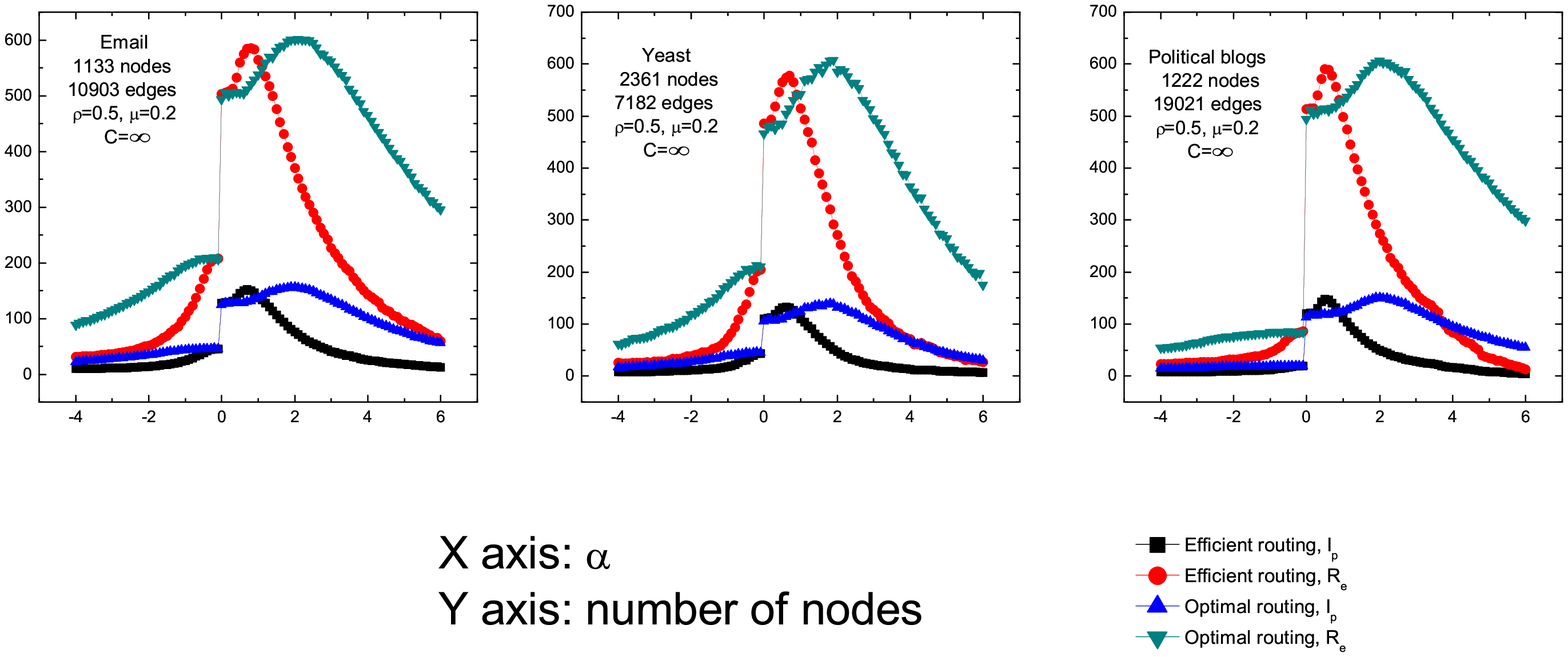}
\caption{$I_p$ and $R_e$  vs.  $\alpha$ for the optimal  routing and the efficient routing on three real-world networks.   The results are the average over 100 independent runs. }
\end{figure}
\section{Conclusion}
In summary, we propose a traffic-driven SIR epidemic model and study the impacts of several factors on our model. We find that the epidemic spreading is greatly affected by the load distribution, and homogeneous load distribution facilitates the epidemic spreading. Increasing the network density or network homogeneity will enhance the epidemic spreading. Large-degree nodes have dual effects on the epidemic spreading, since large-degree infected nodes facilitate the epidemic spreading, while large-degree recovered nodes greatly inhibit the epidemic spreading. To realize an intense and wide epidemic spreading, the paths for the delivery of packets are generally biased towards including small-degree nodes. Increasing packet generation rate generally favors the epidemic spreading. However, when the amount of generated packets is larger than the delivery capacity of the network, there will exit traffic congestion, which blocks the epidemic spreading. Also, we find similar impacts of different static routing protocols on the traffic-driven SIR epidemic spreading. Our work helps understanding the interplay between traffic dynamics and epidemic spreading, and provides some clues for network immunization. 

\ack
This work was  supported by the Natural Science Foundation of China (Grant No. 61304154), the Specialized Research Fund for the Doctoral Program of Higher Education of China  (Grant No. 20133219120032), and the Postdoctoral Science Foundation of China (Grant No. 2013M541673).


\section*{References}
\begin{thebibliography}{10}



\bibitem{Dor08}
S. N. Dorogovtsev,  A. V.  Goltsev, J. F. F.  Mendes, {\it Rev. Mod. Phys.} {\bf 80}    
(2008) 1275.
\bibitem{Barrat}
A. Barrat, M. Barthelemy, A. Vespignani, Dynamical processes on complex networks, Cambridge University Press, Cambridge, 2008.
\bibitem{Newman09}
M. E. J. Newman, Networks: an introduction, Oxford University Press, 2009.
\bibitem{pastor2014}
R. Pastor-Satorras, C. Castellano, P. Van Mieghem, A. Vespignani,  arXiv:1408.2701, 2014.
\bibitem{barabasi09}
A. L. Barab\'{a}si, {\it Science} {\bf 325} (2009) 412. 
\bibitem{pastor01}
R. Pastor-Satorras, A. Vespignani, {\it Phys. Rev. Lett.} {\bf 86} (2001) 3200.
\bibitem{Yangz12}
Z. Yang, T. Zhou, {\it  Phys. Rev. E} {\bf 85} (2012) 056106.
\bibitem{sahneh13}
F. D. Sahneh, C. Scoglio,  P. Van Mieghem, {\it IEEE/ACM Transactions on Networking (TON)} {\bf 21}  (2013) 1609.
\bibitem{MEJNEWMAN02}
M. E. J. Newman, {\it  Phys. Rev. E} {\bf 66} (2002) 016128.
\bibitem{barth04}
M. Barth\'{e}lemy, A. Barrat, R. Pastor-Satorras, A.  Vespignani, {\it Phys. Rev. Lett.} {\bf 92} (2004) 178701.
\bibitem{barth05}
M. Barth\'{e}lemy, A. Barrat, R. Pastor-Satorras, A.  Vespignani, {\it Journal of theoretical biology} {\bf 235} (2005) 275.
\bibitem{karrer10}
B. Karrer, M. E. J. Newman, {\it Phys. Rev. E} {\bf 82} (2010) 016101.
\bibitem{kenah07}
E. Kenah, J.  M.  Robins, {\it Phys. Rev. E} {\bf 76} (2007) 036113.
\bibitem{miller07}
J. Miller, {\it Phys. Rev. E} {\bf 76} (2007) 010101.
\bibitem{goltsev08}
A. V. Goltsev, S. N. Dorogovtsev, J. F. F. Mendes, {\it Phys. Rev. E} {\bf 78} (2008) 051105.
\bibitem{goltsev12}
A. V. Goltsev, S. N. Dorogovtsev, J. G. Oliveira,  J. F. F. Mendes, {\it Phys. Rev. Lett.} {\bf 109} (2012) 128702.
\bibitem{serrano06}
M. A. Serrano, M. Bogu\~{n}\'{a},  {\it Phys. Rev. Lett.} {\bf 97} (2006) 088701.
\bibitem{miler09}
J. Miller, {\it Phys. Rev. E} {\bf 80} (2009) 020901.
\bibitem{gang05}
Y. Gang, Z. Tao, W. Jie, F. Zhong-Qian, W. Bing-Hong, {\it Chinese Physics Letters} {\bf 22} (2005) 510.
\bibitem{chu11}
X. Chu, Z. Zhang, J. Guan, S. Zhou, {\it Physica A} {\bf 390} (2011) 471.
\bibitem{PASTOR02}
R. Pastor-Satorras, A. Vespignani, {\it Phys. Rev. E} {\bf 65} (2002) 036104.
\bibitem{van2012}
P. Van Mieghem, {\it Computer Communications} {\bf 35} (2012) 1494.
\bibitem{starnini}
M. Starnini, A. Machens, C. Cattuto, A. Barrat, R. Pastor-Satorras, {\it Journal of theoretical biology} {\bf 337} (2013) 89.
\bibitem{Granell}
C. Granell, S. G\'{o}mez, A. Arenas,  {\it Phys. Rev. Lett.} {\bf 97} (2013) 128701.
\bibitem{zhao14}
D. W. Zhao, L. H. Wang, S. D. Li, Z. Wang, L. Wang, B. Gao, {\it Plos one} {\bf 9} (2014) e112018.
\bibitem{meloni}
S. Meloni, A. Arenas, Y. Moreno, {\it PNAS} {\bf 106} (2009)  16897.
\bibitem{yang11}
H. X. Yang, W. X. Wang, Y. C. Lai, Y. B. Xie, B. H. Wang, {\it Phys. Rev. E} {\bf 84} (2011) 045101.
\bibitem{yang14}
H. X. Yang, Z. X. Wu, {\it J. Stat. Mech.} {\bf 3} (2014) P03018.
\bibitem{yangh13}
H. X. Yang, Z. X. Wu, B. H. Wang, {\it Phys. Rev. E} {\bf 87} (2013) 064801.
\bibitem{yan06}
G. Yan, T. Zhou, B. Hu, Z. Q. Fu, B. H. Wang, {\it Phys. Rev. E} {\bf 73} (2006) 046108.
\bibitem{arenas01}
A. Arenas, A. D\'{i}az-Guilera, R. Guimer\`{a}, {\it Phys. Rev. Lett.} {\bf 86} (2001) 3196.
\bibitem{Erdos}
P. Erd\"{o}s, A. R\'{e}nyi, {\it Publ. Math. Inst. Hung. Acad. Sci.} {\bf 5} (1960) 17.
\bibitem{Goh01}
K.-I. Goh, B. Kahng, D. Kim, {\it Phys. Rev. Lett.} {\bf 87} (2001) 278701.
\bibitem{guimera}
R. Guimer\`{a}, L. Danon, A. Diaz-Guilera, F. Giralt,  A. Arenas, {\it Phys. Rev. E} {\bf 68} (2003) 065103(R).
\bibitem{shiwei}
S. W. Sun, L. J. Ling, N. Zhang, G. J. Li, R. S. Chen, {\it Nucleic Acids Research} {\bf 31} (2003)  2443.
\bibitem{damic05}
L. A. Adamic, N. Glance,  in Proceedings of the WWW-2005 Workshop on the Weblogging Ecosystem, 2005. 
\bibitem{Wangk11}
K. Wang, Y. Zhang, S. Zhou, W. Pei, S. Wang, T. Li,  {\it Physica A} {\bf 390} (2011) 2593.
\end{thebibliography}
\end{document}